\documentclass[10pt]{iopart}

\usepackage{graphicx}
\expandafter\let\csname equation*\endcsname\relax
\expandafter\let\csname endequation*\endcsname\relax
\usepackage{amsmath}
\usepackage{cite}
\usepackage{textgreek}
\begin{document}


\title{Rapid, broadband spectroscopic temperature measurement of CO$_2$ using VIPA spectroscopy}

\author{Andrew Klose$^{1}$\footnote{Present address:
Department of Chemistry, Augustana University, Sioux Falls, SD, USA.}}
\author{Gabriel Ycas$^1$}
\author{Flavio C.~Cruz$^{1,2}$}
\author{Daniel L.~Maser$^{1,3}$}
\author{Scott A.~Diddams$^{1,3}$}

\address{$^1$ National Institute of Standards and Technology, Boulder, CO, USA 80305 \\
$^2$ Instituto de Fisica Gleb Wataghin, Universidade Estadual de Campinas, Campinas, SP, 13083-859, Brazil \\
$^3$ Department of Physics, University of Colorado, Boulder, CO, USA 80309}
\ead{aklose@augie.edu}


\date{\today}

\begin{abstract}
Time-resolved spectroscopic temperature measurements of a sealed carbon dioxide sample cell were realized with an optical frequency comb combined with a two-dimensional dispersive spectrometer.  A supercontinuum laser source based on an erbium fiber mode-locked laser was employed to generate coherent light around 2000~nm (5000 cm$^{\text{-} 1}$).  The laser was passed through a 12-cm-long cell containing CO$_2$, and the transmitted light was analyzed in a virtually-imaged phased array- (VIPA-) based spectrometer.  Broadband spectra spanning more than 100~cm$^{\text{-1} }$ with a spectral resolution of roughly 0.075~cm$^{\text{-} 1}$ (2.2 GHz) were acquired with an integration period of 2~ms.  The temperature of the CO$_2$ sample was deduced from fitting a modeled spectrum to the line intensities of the experimentally acquired spectrum. Temperature dynamics on the time scale of milliseconds were observed with a temperature resolution of 2.6~K.  The spectroscopically-deduced temperatures agreed with temperatures of the sample cell measured with a thermistor. Potential applications of this technique include quantitative measurement of carbon dioxide concentration and temperature dynamics in gas-phase chemical reactions (e.g., combustion), and plasma diagnostics.  
\end{abstract}


\maketitle

\section{Introduction}
Spectroscopic temperature measurements using coherent or incoherent light sources coupled to optical spectrometers have been accomplished by measuring the distribution amongst the quantized energy levels of a given sample, for example, the ro-vibrational intensities of a molecule \cite{hanson78}.  Moreover, the dependence of the experimental spectrum on temperature and pressure is also observed via line-shape broadening and line-center shifts. The temperature and pressure of the system can be deduced by comparing the experimental spectrum to a modeled spectrum that includes temperature and pressure dependence.  Spectroscopic temperature measurements based on sample absorption have several advantages compared to well-established temperature measurement tools such as thermocouples, thermistors, and blackbody radiation spectral measurements.  For example, in an optical setup, light can be passed through a gas sample to measure the temperature \textit{in situ}, and integrated across a remote path of the gas sample, as compared to a point measurement from a device mounted onto the container wall. 

There has been much research into the development of spectroscopic temperature measurement sensors. Measurements with such systems have been accomplished by rapidly scanning (10~kHz to 1~MHz) single-frequency lasers over absorption features in sample gases in harsh environments to measure the linewidth, centroid, and intensity \cite{sanders02,rieker07,lytkine10}.  One disadvantage of previous spectroscopic temperature measurements is the tradeoff of spectral bandwidth for rapid scanning on time scales of under 1~ms.  Specifically, using a single diode laser source, it has been difficult to achieve rapid spectroscopic measurement with spectral bandwidth of more than 15~cm$^{-1}$.  However, in cases where sample absorption is weak, or where the sample is contained in an interfering matrix, it is advantageous to interrogate more sample absorption features by measuring over a larger optical bandwidth.  Previously, Fourier-transform techniques have been used with much success for broadband measurements to monitor dynamic systems \cite{woodbridge88}. To achieve the desired time resolution in this case, the system under study must be readily reproducible and the spectrometer must be synced with ``time-zero'' of the event being studied \cite{woodbridge88}. Each desired Fourier stage position must be measured as a function of event time delay.  

Here, the implementation of an optical frequency comb coupled to a two-dimensional spatially-dispersive spectrometer has been realized to measure temperature dynamics on time scales less than 10~ms in a single shot measurement.  Individual spectra spanning more than 100~cm$^{\text{-} 1}$ with a spectral resolution of 0.075~cm$^{\text{-} 1}$ (2.2 GHz) were acquired with an integration period of 2~ms. The time between consecutive spectra was 8.3~ms, limited by the 120~Hz frame rate of the camera integrated into the spectrometer. Previously, a similar system was used to measure time-dependent concentrations of transient species in photo-induced chemical reactions involving free-radicals \cite{fleisher14}.  In the present work, a spectroscopic temperature measurement technique is demonstrated by observing ro-vibrational intensity changes in absorption spectra of a CO$_2$ sample during resistive heating of a sample cell.  The $20012 \leftarrow 00001$ spectral band of CO$_2$ centered at 4977.830~cm$^{-1}$ \cite{rothman2009} was probed in the current measurements.

\section{Experimental Setup}
\begin{figure}
\centering
\includegraphics{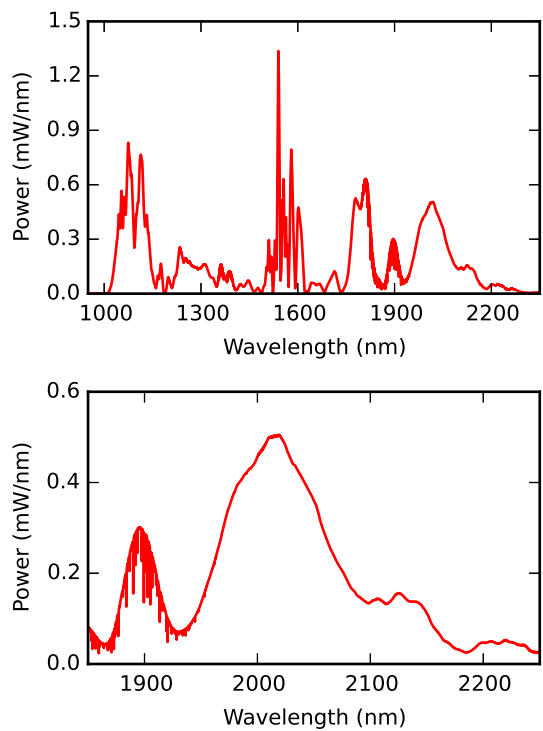}
\caption{Example output supercontinuum spectrum from the Er mode-locked laser.  The full spectrum is shown in the top panel, and an expanded view of the 2000~nm spectral region is presented in the bottom panel. Depletions near 1900~nm are due to water absorption in the spectrometer used for the measurement.}
\label{fig:source-spectrum}
\end{figure}

\begin{figure}
\centering
\includegraphics{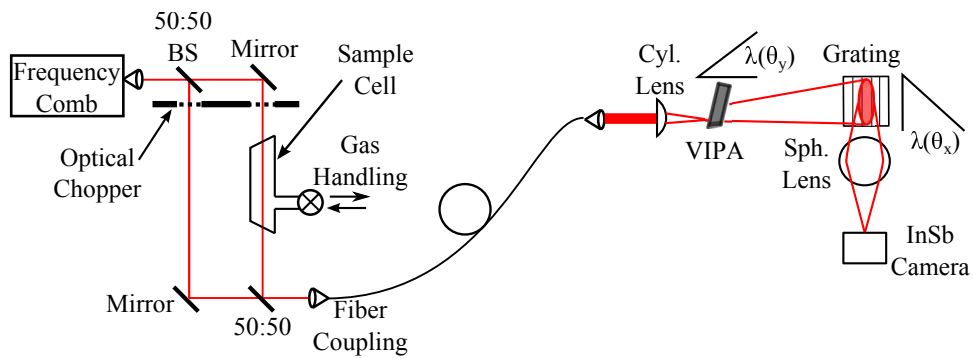}
\caption{Schematic diagram of the experimental setup presented in the current work.  Light from a mode-locked laser was split through two arms of the gas-cell measurement station, combined, and coupled into single mode fiber.  The fiber-coupled light was delivered to the VIPA spectrometer, which is described in detail in the text.  The reference signal from the optical chopper, which chopped both arms of the gas-cell measurement station, was used to sync the InSb camera in the spectrometer setup.  The camera operated at 120~Hz and recorded interleaving images from each arm of the sampling setup.}
\label{fig:exp-setup}
\end{figure}

The laser light used for spectroscopy was produced via supercontinuum generation from the amplified output of a 250~MHz erbium fiber mode-locked laser.  The performance of the laser system is described in detail elsewhere \cite{klose14}.  Here the system is briefly described. The output of the erbium oscillator was amplified and compressed in fiber to achieve 70~fs full-width at half maximum (FWHM) pulses with an average power of 350~mW and an optical spectrum that spanned roughly 1500~nm to 1600~nm.  The femtosecond pulses were delivered to a highly nonlinear fiber in which the optical spectrum was broadened to a bandwidth that spanned from 1000~nm to 2200~nm.  The long-wavelength portion of the optical spectrum from 1980~nm to 2050~nm (4880~cm$^{-1}$ to 5050~cm$^{-1}$) was used for spectroscopy of CO$_2$.  A representative optical spectrum from the laser system is presented in Figure~\ref{fig:source-spectrum}.  The fiber-coupled output light was directed though a sample followed by a spectrometer system shown schematically in Figure~\ref{fig:exp-setup}.  The laser light was launched into free space and split into two paths using a 50:50 beam splitter.  One path passed through a temperature controlled 12-cm-long sample cell that had uncoated ZnSe windows positioned at Brewster's angle relative to the incident laser light.  The second optical path was a reference path that did not pass through the sample cell.  The beams from each path were recombined, coupled into an optical fiber, and delivered to a virtually-imaged phased array- (VIPA-) based spectrometer \cite{shirasaki96,diddams07}.

The fiber-coupled input to the spectrometer was launched into free-space and focused into a 1-mm-thick VIPA etalon (free spectral range (FSR)~$\sim$~51~GHz) using a cylindrical lens ($f = 50$~mm).  The VIPA was tilted roughly 4 degrees from normal to the incident light.  The highly-dispersed output from the VIPA had overlapping orders separated by the VIPA FSR. The light was impinged onto a reflective diffraction grating orientated orthogonal to the VIPA dispersion to separate the overlapping VIPA orders.  The two-dimensionally dispersed light was imaged onto a InSb camera using an $f=250$~mm spherical lens; roughly 1 pW per comb mode was incident on the camera.  The InSb camera had an array of $640 \times 512$ pixels with a pixel pitch of 20~\textmu m, had a maximum full frame rate of 120~Hz, and was cooled to 77~K using liquid nitrogen.

The InSb camera was triggered by an optical chopper that acted on both the sample beam path and reference beam path.  The chopped light from the sample path was out of phase with the reference path.  The trigger signal allowed for the sample and reference images to be interleaved during data acquisition. Such interleaving allowed for rapid normalization of each acquired spectrum and minimized fast time-dependent drifts of the experimental setup.  A second reference was acquired when the gas cell was under vacuum.  The vacuum-cell reference allowed for calibration of the static wavelength dependent difference (from, e.g., etalons) between the two beam paths in the sampling setup.

\section{Results and Discussion}
\begin{figure}
\centering
\includegraphics{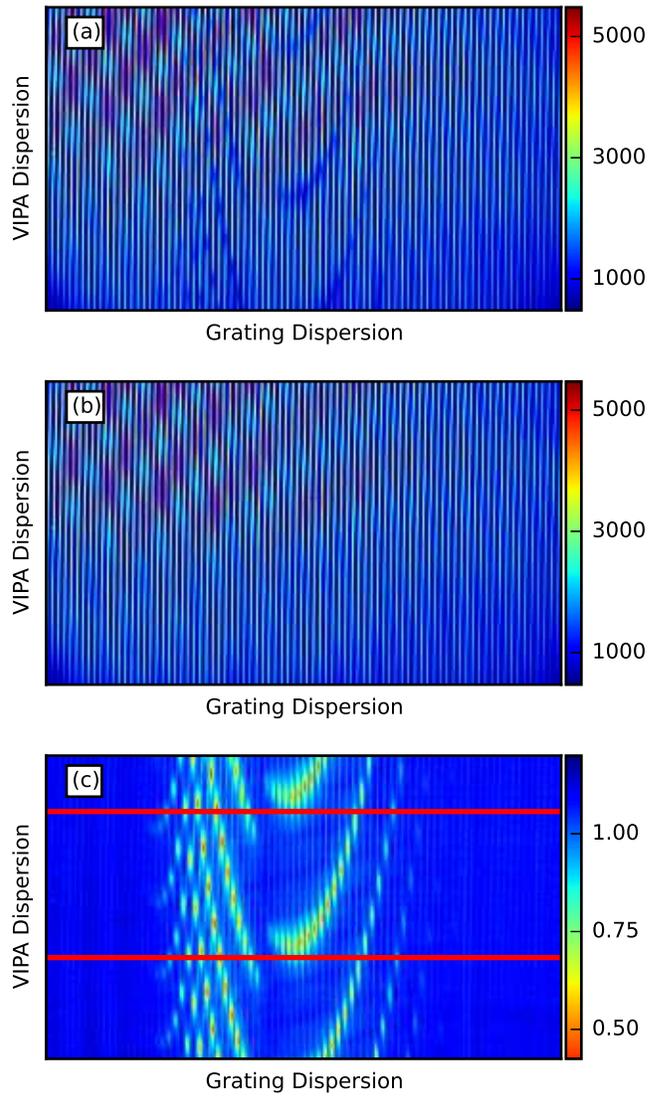}
\caption{(a) Raw camera image of light passing through the CO$_2$-filled sample cell.~(b) Raw camera image of light passing through the reference arm.  Images presented in (a) and (b) were taken in consecutive camera frames.  (c) Image of CO$_2$ absorption after image normalization described in the text.  Red lines indicate the free spectral range of the VIPA.}
\label{fig:image-sequence}
\end{figure}

Consecutive camera images of light passing through the CO$_2$-filled sample cell and reference arm are presented in panels (a) and (b) of Figure~\ref{fig:image-sequence}, respectively.  An image of the absorption of CO$_2$ was generated by subtracting the dark image from all other images, then dividing the CO$_2$ sample image by the consecutive reference arm measurement. A similarly divided image was obtained with the cell under vacuum.  The images were multiplied together as
\begin{equation}
\frac{I_{\text{cell,CO2}}}{I_{\text{ref,CO2}}} \times \frac{I_{\text{ref,vac}}}{I_{\text{cell,vac}}},
\end{equation}
where $I$ denotes the image taken from the beam passing through the sample cell (cell) or reference path (ref) with CO$_2$ in the cell (CO2), or the cell under vacuum (vac). The two-step calibration minimized both short- and long-term wavelength-dependent drifts in the spectrometer.  All images were acquired with a camera integration period of 2~ms per image. The CO$_2$ absorption is shown in panel (c) of Figure~\ref{fig:image-sequence}.  The vertical stripes are due to the VIPA dispersion, and the free spectral range of the etalon was empirically determined by inspection of the camera image.  The VIPA FSR is depicted as horizontal red lines in Figure~\ref{fig:image-sequence}. A typical transmission spectrum was constructed by appending consecutive stripes together to obtain a one-dimensional wavelength axis.  The pixel intensity along each stripe is indicative of sample transmission.

The resolution of the spectrometer was measured to be ($0.075 \pm 0.003$)~cm$^{-1}$, or ($2.2 \pm 0.1$)~GHz, and was determined by launching a tunable single frequency external cavity diode laser (ECDL) into the spectrometer.  The diode laser was scanned from 1980~nm to 2050~nm (4880~cm$^{-1}$ to 5050~cm$^{-1}$) and camera images were recorded throughout the wavelength range.  The resolution was consistent within 0.1~GHz throughout the portion of the image used to construct the line out spectrum.  The limiting factor of the spectral resolution was the optical coatings on the VIPA etalon.  The VIPA used in the present work was employed in previous measurements \cite{cossel10}, and reflectance of the VIPA surfaces decreased above 2000~nm, limiting the finesse and ultimate resolution. However, the spectrometer presented here provided ample means to measure the absorption features of CO$_2$, which had broadened widths of more than 5~GHz.

\begin{figure}
\centering
\includegraphics{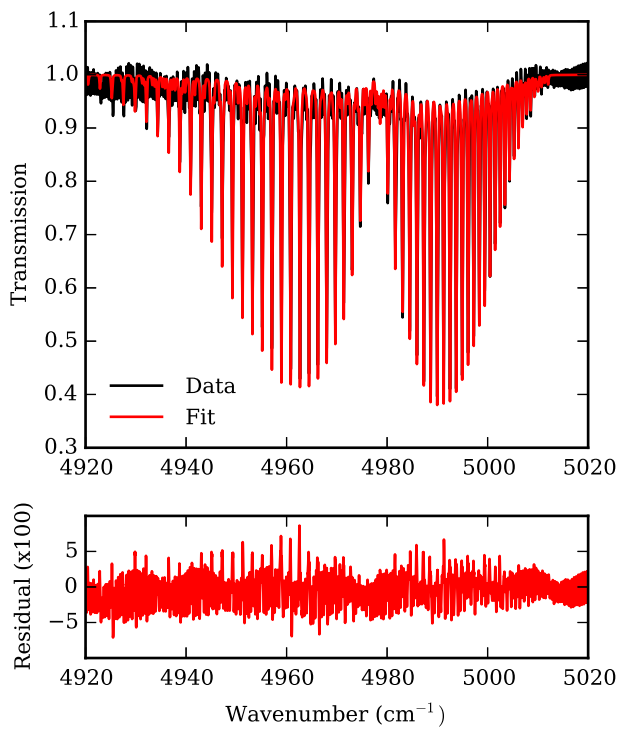}
\caption{Fit result of a spectrum recorded when the measured temperature of the CO$_2$ cell was 328~K. The data and modeled spectrum are displayed on the upper panel, and the residuals are presented in the lower panel. The deduced temperature from the fit was 327~K.}
\label{fig:fit-ex}
\end{figure}

Calibration of the frequency axis of the spectrum was done by fitting the pixel position of absorption peaks in the experimental spectrum to the peak positions of a modeled absorption spectrum obtained from the High Resolution Transmission (HITRAN) database \cite{rothman2009}.  A third-order polynomial fit of the experimental peak position to the modeled spectrum provided a hertz-per-pixel calibration of the experimental spectrum.  After calibration, a least-squares fitting routine was used to achieve optimum agreement between the modeled and experimental spectra.  In the minimization routine, the modeled spectrum, which is detailed in the appendix of Ref.~\cite{rothman98}, was allowed to vary in temperature and pressure.  The path length through the CO$_2$ sample was fixed to 12~cm. The modeled absorbance spectrum was determined by calculating the line intensity from the number density and interaction length of the sample.  A Voigt line shape was used in the model.  The line shape accounted for pressure and temperature broadening, pressure shift of the line centroids, and temperature dependence of the line intensities \cite{rothman98}.  While both temperature broadening and line intensities were simultaneously accounted for in the routine, the most sensitive effect for deducing the temperature from the fit was the dependence of each line intensity on temperature; the modest resolution of the present apparatus limited the observable effect of temperature broadening in the experimental spectra.  Additionally, given the present resolution, line mixing and speed dependence of the line shape were not included in the model.  The modeled spectrum was convolved with the instrument response, which was empirically determined from ECDL measurements to be well-approximated by a Gaussian function with a 2.2~GHz FWHM.  A Lorentzian instrument response function, with a FWHM of 2.2~GHz, was used as response function as well.  The deduced temperature from the independent fits with the differing instrument response functions were consistent within the temperature uncertainty of 2.6~K, which is discussed later in the paper.   After minimization, the frequency axis was re-calibrated using a third-order polynomial fit of the peaks in the experimental spectrum to peak positions of the optimum modeled spectrum obtained in the minimization routine.  The alternating frequency calibration of the experimental spectrum and least-squares fitting of the modeled spectrum continued until the optimum modeled spectrum converged, which typically occurred after two iterations.  A fit result is presented in Figure~\ref{fig:fit-ex}. Prior to measurement of the spectrum shown in Figure~\ref{fig:fit-ex}, the sample cell was measured to have a pressure of $(700 \pm 50)$~Torr and a temperature of $(328 \pm 3)$~K. The pressure and temperature deduced from the fit were 722~Torr and 327~K, respectively.  The large uncertainty in measured pressure was a result of the use of an uncalibrated analog pressure transducer.  Additionally, the transducer was mounted near, but not adjoining, the laser-interaction region, and the transducer was not at the same temperature as the interaction region.  The pressures deduced from the fits agreed with the expected behavior as a function of increasing temperature (i.e. the ideal gas law), but further analysis of the uncertainty in the pressures deduced from the fits was not possible given the lack of a reliable truth measurement.

\begin{figure}
\centering
\includegraphics{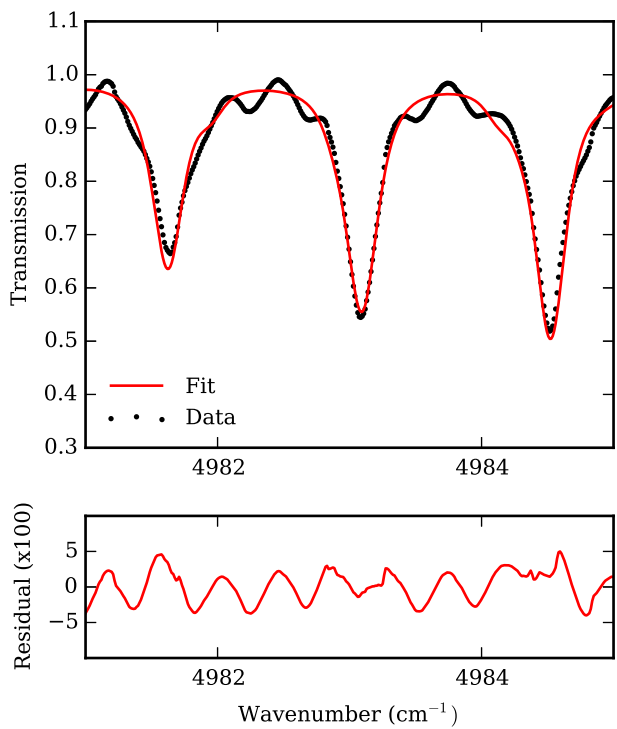}
\caption{Zoomed in portion of a fit of the spectrum.  The black dots indicate each vertical pixel moving along the vertical stripes from the VIPA image.}
\label{fig:fit-ex-zoom}
\end{figure}

A zoomed in portion of the fit result is shown in Figure~\ref{fig:fit-ex-zoom}. The black dots in the figure indicate the signal from each pixel of the vertical stripes from the VIPA image.  There was an etalon visible in the residual of the fit, and this effect was the limiting factor for the sensitivity of the deduced temperature in the present measurement.  During the course of data collection, the etalon was minimized as much as possible. Because the goal in the present work was to fit spectra recorded with a single image, averaging multiple spectra to reduce the etaloning was not done.  On the other hand, while analyzing the data, fitting, and alternatively Fourier filtering, of the etalon was attempted.  However, these techniques were unsuccessful in removing the etalon because of small, but non-zero, non-linearities in the camera image-to-lineout spectrum conversion.  Because of the non-linearities, the etalon did not have a single frequency and could not be removed or modeled successfully. In the future, if one wanted to increase the sensitivity of the measurement, fitting out, or better yet intrinsically reducing, the etalons would be optimal.

\begin{figure}
\centering
\includegraphics{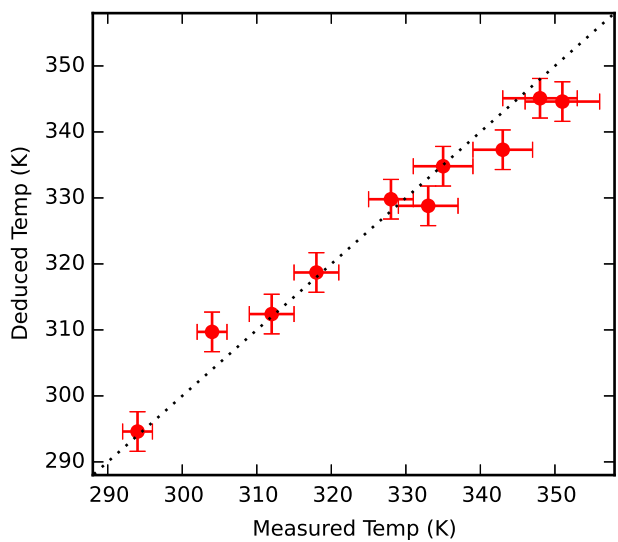}
\caption{Comparison of spectroscopically-deduced (y axis) and thermistor-measured (x axis) temperatures over a 50~K range.  The dotted black line is the curve $y=x$, not a fit.}
\label{fig:fit-temp}
\end{figure}

The spectrum presented in Figure~\ref{fig:fit-ex} was one of a series of measurements where the cell was resistively heated and allowed to come to an equilibrium temperature. Spectra were acquired as a function of measured cell temperature.  These spectra were fitted with modeled spectra \cite{rothman98}, and the temperature of the gas was deduced from each fit.  The deduced temperatures for each fit are plotted against the temperature measured with a thermistor in Figure~\ref{fig:fit-temp}.  Uncertainty in the thermistor-measured temperatures was due to gradients present from uneven heating of the 12-cm-long cell.  The reproducibility of the thermistor temperature measurements for a given heating set point was better than 1~K.  Nonetheless, the deduced temperatures agree with the thermistor measurements over a range of over 50~K.

\begin{figure}
\centering
\includegraphics{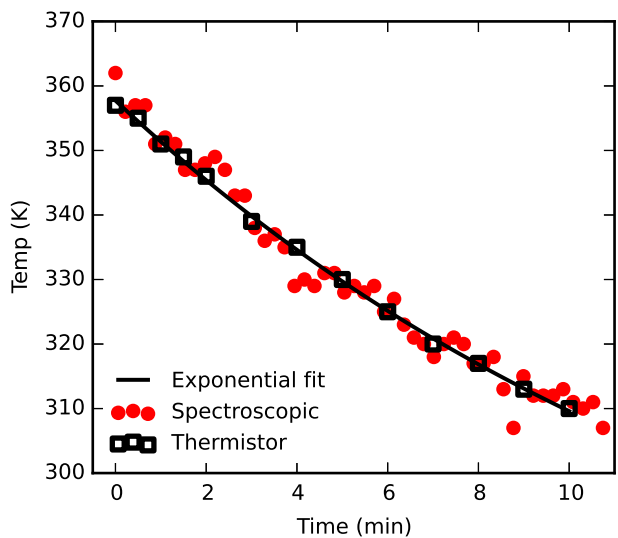}
\caption{Temperature decrease of the CO$_2$ sample cell.  The heating element was turned off at time zero of the plot.  The black squares are thermistor measurements, and the red circles are spectroscopically-deduced temperatures.  The solid black line is an exponential fit to the thermistor temperature measurements.}
\label{fig:T-dynamic}
\end{figure}

\begin{figure}
\centering
\includegraphics{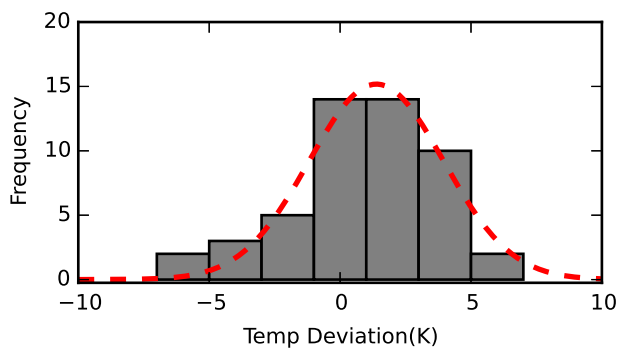}
\caption{Histogram of the average deviation of spectroscopically-deduced temperature from the exponential fit of the temperature decay of the CO$_2$ cell presented in Figure~\ref{fig:T-dynamic}.  The dashed line is the fit of a Gaussian function to the histogram.  The mean and FWHM of the Gaussian function were 1.4~K and 2.6~K, respectively. }
\label{fig:histo}
\end{figure}

Dynamic temperature changes were investigated by resistively heating the cell to roughly 360~K, turning off the heating element, and recording CO$_2$ spectra over the course of ten minutes.  Spectra were acquired every 13~s, with a camera integration period of 2~ms for each measurement.  The temperature of the sample as a function of time was deduced by fitting each spectrum to the model described previously.  The result of the time-dependent spectroscopic temperature measurements is presented as red circles in Figure~\ref{fig:T-dynamic}.  The open black squares represent temperature measurements of the cell using the thermistor. The temperature measurements, deduced from spectra that were acquired in 2~ms, agree with the thermistor-measured temperature over the 10~min time scale of the dynamic measurement.  The thermistor-measured temperatures were fit with an exponential decay function; the result is plotted as a black line on Figure~\ref{fig:T-dynamic}.  The deviation of the spectroscopically-deduced temperatures from the exponential fit is presented as a histogram in Figure~\ref{fig:histo}.  The histogram was fitted with a Gaussian function, and the mean and FWHM of the fitted distribution were 1.4~K and 2.6~K, respectively. The 1.4~K accuracy of the deduced temperatures compared to exponential temperature decay fit is within the uncertainty of the sample-cell temperature.    

\begin{figure}
\centering
\includegraphics{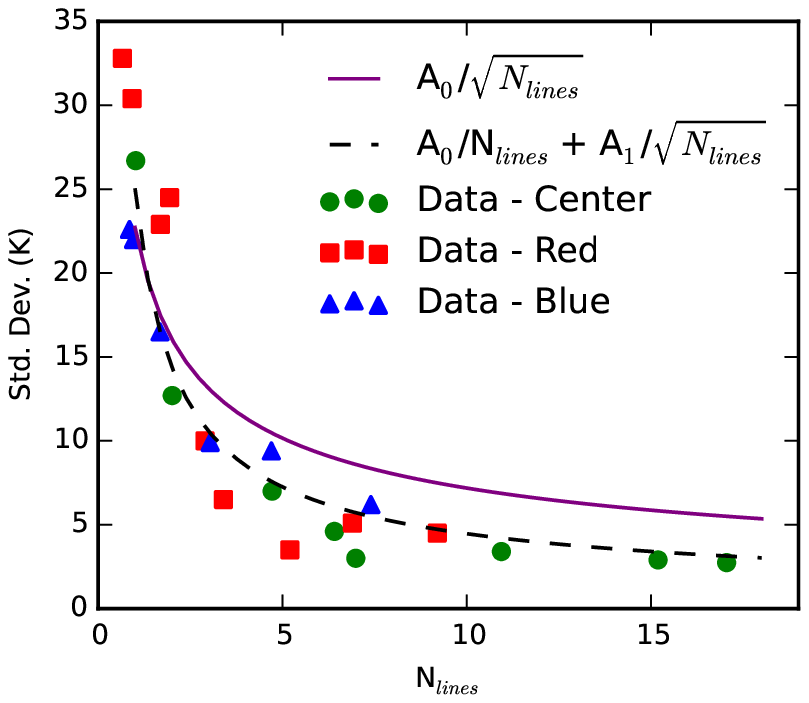}
\caption{Standard deviation of spectroscopically-deduced temperatures as a function of the number of spectral features measured.  $N_{\text{ lines}}$denotes a normalized quantity that is the effective number of absorption lines that attenuate nearly 100~\% of the incident light.  The independent data sets are described in the text.  The solid purple and dashed black curves are fits of the functions $\left( A_0 / \sqrt{N_{\text{ lines}}} \right)$ and $\left( A_0 / N_{\text{ lines}} + A_1 / \sqrt{N_{\text{ lines}}} \right)$ to the data, respectively.}
\label{fig:T-unc}
\end{figure}

More notably, the fit result in Figure~\ref{fig:histo} indicates that the uncertainty in the spectroscopically-deduced temperatures was 2.6~K. The change in uncertainty, or increase in sensitivity, was investigated here by deducing the sample temperature by fitting sub-sections of the full measured bandwidth.  The 50 spectra acquired during the sample temperature decay were independently fitted multiple times with differing numbers of absorption features in the spectrum.  A total of 53 absorption features were visible above the background noise.  The differing intensity of the absorption features was normalized by adding the maximum absorption of all of the peaks included in each fit.  This sum gives an equivalent number of peaks, denoted $N_{\text{ lines}}$, that absorb nearly 100~\% of the incident light.  A related approach using the integrated area peaks was considered in the case of gas mixing in the supplementary information of Ref~\cite{adler10a}.  In the present work, the standard deviation of the difference between the deduced temperatures and exponential decay fit are plotted as a function of the number of normalized absorption lines in Figure~\ref{fig:T-unc}. The spectra were fitted in three separate sequences.  First, the center of the spectrum was fitted, and additional peaks on each side were systematically included as the fitted spectral window was widened.  The result from this sequence is presented as green circles in Figure~\ref{fig:T-unc}.  Second, the far red portion of the spectrum, near 4920~cm$^{-1}$, was fitted, and additional peaks with centroids at higher wavenumbers were systematically included as the fitting window was widened.  Those data are presented as red squares in Figure~\ref{fig:T-unc}.  Third, the far blue portion of the spectrum, near 5000~cm$^{-1}$, was fitted, and additional peaks with centroids at lower wavenumbers were systematically included as the fitting window was widened; the result is presented as blue triangles in Figure~\ref{fig:T-unc}. The behavior, regardless of the fitting window was consistent. The deviation is large, near 25~K for low values of $N_{\text{ lines}}$, and saturates near 3~K for large values of $N_{\text{ lines}}$.  The saturation behavior is an indication that the temperature gradient across the cell may be a limiting factor in the sensitivity. Nonetheless, a temperature sensitivity of $\pm 2.6$~K was achieved in the present case of a sample with roughly 50 peaks absorbing between 5~\% and 60~\% of the incident light.  If the discrepancies of each peak intensity were random, it would be expected that the uncertainty would decrease as the square root of independent absorption features of equal intensity.  The function $A_0 / \sqrt{N_{\text{ lines}}}$, where $A_0$ is a free parameter, was fitted to these data and is shown as a solid purple line in Figure~\ref{fig:T-unc}. However, in the present case, the absorption for the CO$_2$ peaks are not independent, but rather are coupled for a given temperature.  Thus, the uncertainty of the spectroscopically deduced temperature decreases faster as a function of equivalent peaks than the square-root function. The correlation can be modeled by adding in a term with $ 1 / (N_{\text{ lines}})$ dependence, namely by fitting the function $A_0 / (N_{\text{ lines}}) + A_1 / \sqrt{N_{\text{ lines}}}$, where $A_0$ and $A_1$ are free parameters, to the data.  The fit of the two-term function, shown as the dashed black line in Figure~\ref{fig:T-unc}, reproduces the trend in the data, and illustrates the correlation of fitting multiple spectral feature intensities.  This result highlights another advantage of the simultaneous measurement of the full ro-vibrational band.

\section{Conclusions}
Rapid spectroscopic temperature measurements of a gaseous carbon dioxide sample were realized via absorption spectroscopy. An Er-based mode-locked laser was used to generate a supercontinuum spectrum, and light near 2000~nm (5000~cm$^{-1}$) was used to interrogate the CO$_2$ sample.  The light was delivered to a VIPA-based two-dimensional dispersive spectrometer, which provided the means to record a sample transmission spectrum with more than 100~cm$^{-1}$ bandwidth simultaneously with 2~ms of integration time.  Deduced temperatures from line intensities allowed for the determination of the CO$_2$ sample temperature with an uncertainty of $\pm 2.6$~K.  Temperature dynamics on a time scale of ten minutes were measured with the present system, and agreed with thermistor temperature measurements of the sample cell.  It is intended to implement the present spectrometer to measure temperature dynamics of CO$_2$ in systems, such as photo-induced reactions, that change on time scales of milliseconds.  It is possible to increase the frame rate of the InSb camera by reducing the image size acquired.  This trade-off of faster acquisition may outweigh the reduction in measured optical bandwidth for some applications. Other frequency comb-based spectroscopic techniques, that do not require a liquied nitrogen-cooled InSb camera, may be advantageous for rapid spectroscopic temperature measurements as well.  For example, implementation of a dual-comb spectroscopic setup \cite{keilmann04, coddington08, giaccari08, rieker14} would allow for time resolution of less than 1~ms, but may require averaging of multiple shots.  The implementation of these techniques will allow for rapid and broadband spectroscopic temperature measurements of dynamic systems.

\ack
This work was supported in part by the NIST Greenhouse Gas and Climate Science Measurements Program. The authors thank T.A.~Johnson, L.~Nugent-Glandorf, F.~Quinlan, and F.~Giorgetta for useful discussions, J.~Ye for providing the VIPA etalon, and M.~Hirano of Sumitomo for providing the nonlinear fiber.  A.K.\ acknowledges support from the National Research Council Postdoctoral Fellowship Program, and F.C.C.\ acknowledges support from CNPq, Fapesp and NIST.  This work is a contribution of the United States Government and is not subject to copyright in the United States.

\section*{References}
\bibliographystyle{unsrt}
\bibliography{refs}


\end{document}